\documentclass[conference]{IEEEtran}

\usepackage{amssymb}
\usepackage{ifpdf}
\usepackage{cite}

\ifCLASSINFOpdf
  \usepackage[pdftex]{graphicx}
  \DeclareGraphicsExtensions{.pdf,.jpeg,.png}
\else
  \usepackage[dvips]{graphicx}
  \DeclareGraphicsExtensions{.eps}
\fi

\usepackage[cmex10]{amsmath}
\usepackage{algorithmic}
\usepackage{algorithm}
\usepackage{array}
\usepackage{url}
\usepackage{fancyhdr}
\begin{document}
  \setlength{\parskip}{1mm}
  \setlength{\parindent}{5mm}
  \setlength{\partopsep}{0mm}

\title{Joint Routing, Scheduling And Power Control For Multihop Wireless Networks \\ With  Multiple Antennas}

\author{\IEEEauthorblockN{Harish Vangala}
\IEEEauthorblockA{ECE Department,\\
Indian Institute of Science,\\
Bangalore, India.
}
\and
\IEEEauthorblockN{Rahul Meshram}
\IEEEauthorblockA{ECE Department,\\
Indian Institute of Science,\\
Bangalore, India.
}
\and
\IEEEauthorblockN{Prof. Vinod Sharma}
\IEEEauthorblockA{ECE Department,\\
Indian Institute of Science,\\
Bangalore, India.
}
}

\maketitle
\begin{abstract}
We consider the problem of Joint Routing, Scheduling and Power-control (JRSP) problem for multihop wireless networks (MHWN) with multiple antennas. We extend the problem and a (sub-optimal) heuristic solution method for JRSP in MHWN with single antennas. We present an iterative scheme to calculate link capacities(achievable rates) in the interference environment of the network using SINR model. We then present the algorithm for solving the JRSP problem. This completes a feasible system model for MHWN when nodes have multiple antennas. We show that the gain we achieve by using multiple antennas in the network is \emph{linear} both in optimal performance as well as heuristic algorithmic performance.

\end{abstract}

\section{Introduction}
Multihop wireless networks are essential for ubiquitous computation and communication. Currently there are many experimental setups of multihop wireless networks around the world. Ad-hoc wireless networks and sensor networks are also examples of multihop wireless networks where it is necessary to employ multiple wireless hops for even the connectivity of the nodes deployed in a particular area. However, multiple wireless hops pose many new challenges in a network design. But recent studies have shown that some of these challenges can be converted into opportunities by careful network design. Thus new communication paradigms, e.g., opportunistic scheduling, cooperative communication, network coding and multiple antennas have been developed in recent years. Exploiting these techniques together in a multihop setup to optimize the system performance is very challenging. In this paper we will concentrate on designing multihop networks with multiple antennas at each node.

In multihop wireless networks, unlike in the wire-line networks, even the concept of a link between two nodes is not properly defined. Shadowing can, irrespective of the distance between two nodes, cause the channel to be very bad. Even when there is no shadowing, we may or may not be able to communicate directly between two nodes depending on the transmit power used and also if there are other users transmitting in the neighborhood.  Thus, topology of the network, transmit power, link scheduling and routing are all interrelated and for optimal performance one may need to jointly optimize power, scheduling and routing. This problem is computationally intractable and not scalable even for a centralized algorithm \cite{Cao_etal}, \cite{VJ_etal}, \cite{Lin_Shroff}.

Employing multiple antennas at a transmitter and/or at a receiver can provide transmit diversity, receive diversity, increase the capacity of the link and reduce BER. Thus, in wireless networks where bandwidth is scarce, it is important to employ multiple antennas wherever feasible. This increases the degrees of freedom one can exploit to improve the system performance. Thus even for multihop network it is desirable to have multiple antennas at different nodes. However, as mentioned above, even with single antennas, jointly optimizing power, routing and scheduling in a multihop wireless network is very computationally intensive, with multiple antennas the problem becomes extremely challenging. At the moment these are very few studies available for this system. We address this problem in this report.
\section{Related Work}
In wireless networks, it is well known that the traditional layers of the communication network cannot be considered in isolation. Many authors have proposed joint approaches to issues such as power control, scheduling, and routing ~\cite{Cao_etal}, ~\cite{VJ_etal}, ~\cite{Lin_Shroff}, ~\cite{Srikanth_etal}. But there has been little effort in joint optimization of power, link scheduling  and routing. One of the early papers in this direction is ~\cite{Cao_etal}. In ~\cite{VJ_etal}, the techniques of ~\cite{Cao_etal} are applied to a different scenario where the nodes use Energy Harvesting Sensors. Both ~\cite{Cao_etal} and ~\cite{VJ_etal} use \emph{Linear Programming} in their system models by using the technique of Column Generation.

A multihop wireless network with single antennas is studied and solved in a similar way claiming to achieve the solutions for much larger networks than what has been possible thus far in ~\cite{Rosenberg_etal}.

\section{System Model And The Problem Formulation}
We have a network of \emph{half duplex} nodes spread out in space communicating with each other using the common wireless medium. Since the network uses a wireless channel of broadcasting nature, ideally every node can hear the transmissions from every other node. But in practice, a node usually will have a coverage area hence being allowed to transmit to or receive from a limited number of other nodes. This effectively forms a Graph and is the starting point for the problem. But it can be noted that interference continues to exist between any pair of nodes, irrespective of the graph representation. Further, the nodes may be restricted in terms of transmission or reception towards other nodes(neighbors) in the network. This is required for a tractable model in terms of formulating and solving the problem. We term it as \emph{transmission model} of the system.

In the given network, a subset of nodes called \emph{sources} would like to transmit to their  \emph{destinations}, which is another subset of nodes. It in general needs multiple wireless hops to reach the destination. Our objective is to provide a \emph{fair} (to be defined later) data transmission rate to all those \emph{sources} via multiple intermediate nodes.

\subsection{System Model}
We have a network being represented as a fully connected directed graph with no self-loops, denoted as $\mathcal{G(N,L)}$. Here, $\mathcal{N}=\{1,2,\ldots,N\}$ is the set of nodes and $\mathcal{L}=\{1,2,\ldots,L\}$ is the set of directed links in the network. The set of source and destination pairs and the transmissions among them will be called as \emph{flows}. The set of flows is denoted by $\mathcal{F}={1,2,\ldots,F}$. Each flow $f\in\mathcal{F}$ will be associated with a source node and a destination node denoted by $src(f)$ and $dest(f)$. It is allowed to have common source and/or destination nodes among flows. Each $f\in\mathcal{F}$ will also have a set of rate demands denoted by $\{d_i, i\in \mathcal{F}\}$, while the system may decide to allow rates equal to $\{r_i, i\in \mathcal{F}\}$, depending on the availability of network resources. The rates allowed may be lesser or greater or equal to the demanded rates. We define \emph{fairness} as the least fraction in ${\left\{\frac{r_i}{d_i}\right\}}_{i\in \mathcal{F}}$ and denote it by $\lambda$.

All the nodes use multiple but equal number of antennas $a$ for transmission or reception. The total power consumption of a node for transmission can only be a value chosen from a set of $K$ powerlevels (including zero), in any time slot. We restrict all the antennas to be used towards a single node at any time instant. That means, we use point to point transmission model or the beam forming model of transmission in the network. Hence, a node involves in at most one active link, at any point of time. 

Though this model is a simple extension of the model used in \cite{Cao_etal} and \cite{VJ_etal} for single antennas, it is not obvious to formulate and solve the problem for multiple antennas. Further, one cannot simply state the MIMO gain in the final throughput, just by observing that each link gains in its capacity, because of multiple hops being involved in transmissions.

Given the transmission model of the network, not every subset of $\mathcal{L}$ can be activated simultaneously. We define a \emph{mode}, as a valid subset of links being activated simultaneously along with the amounts of powers used on the links. We represent it by an $L\times 1$ vector of powers used on each link. A set of \emph{modes} given a schedule, achieves the link scheduling and routing simultaneously. If such scheduling takes the power availability on an average at each node into account, it becomes \emph{a} solution to the JRSP problem.

The fading channel gains are assumed to be known during the analysis being a centralized setup and are given by set of $a\times a$ matrices $H_{ij} (\forall i,j\in\mathcal{N})$, which represent the channel matrices between nodes $i$ and $j$. We also use the notation $H_{lm}' (\forall l,m\in\mathcal{L})$ to denote the channel between links $l,m$ meaning, the channel between source node of link $l$ and destination node of link $m$. Note that $H_{lm}'=H_{ij}$, where $i=src(l)$ and $j=dest(m)$. Each scalar in the matrices is taken as a circularly Gaussian complex random variable. In \cite{Cao_etal} and \cite{VJ_etal}, these were calculated by using a simple path loss model for a set of nodes randomly spread out over an area, based on the euclidean distances between them. 

Given the power spent on a link (in any \emph{mode}), we retain the distribution of powers over the set of antennas at nodes as a degree of freedom for nodes (and hence is a variable during link capacity calculations). Hence they use \emph{waterfilling}, whenever required. The way we compute link capacities is presented after the mathematical statement of the problem.

\subsection{The Mathematical Formulation}

\begin{subequations}  \label{The_LPP}
  \begin{align} 
      \max \; \lambda &\triangleq {\left(\min_{i}{\left\{\frac{r_i}{d_i}\right\}}\right)} \\\tag{fairness objective}\\
      \text{subject to}: \nonumber \\ \vspace{3mm}
      A X  ~&~ = ~ \mathbf{r} ~, \\\tag{Flow Conservation} \\ 
      X\cdot \underline{1} ~&~ \leq ~  C\cdot \underline{\alpha} ~, \\\tag{Avg. Link Capacities}\\ 
      P\cdot \underline{\alpha} \quad\!  ~&~ \leq ~ \quad\! \underline{P}^{avg} ~, \\\tag{Avg. Power Control} \\
      \underline{\alpha}\cdot \underline{1}  ~&~ = ~ 1 ~, \\\tag{Consistency of scheduling} \\
      X_l^f\geq 0,~\forall f&\in\mathcal{F},~ \forall l\in \mathcal{L} \nonumber \\
      P_n^m \geq 0,~\alpha_m \geq 0,~~ &\forall n\in\mathcal{N},~\forall m\in \mathcal{M} \\\tag{non-negativity of variables}. 
  \end{align}
\end{subequations}

The problem setup used in \cite{Cao_etal} and \cite{VJ_etal} is quite general and the problem formulation under multiple antennas can be made similar. Except that the terms used in it come from a different network scenario in multiple antennas. Its a conventional \emph{Multicommodity Flow problem} from network theory and operations research literature, which is modified and exploited to suit the JRSP problem. The problem's objective is to find the solution to the JRSP problem, that gives maximum \emph{fairness} among the users.
The mathematical statement of JRSP problem contains the well known flow conservation constraints, link capacity constraints and average power constraints. This forms a simple \emph{linear programming problem} (LPP) as given in \eqref{The_LPP}.

In the problem statement \eqref{The_LPP}, '$A$' refers to the node versus link representation of the graph which takes the form of an $N\times L$ matrix with elements $a_{ij}$ taking signed binary values ${0,\pm 1}$, as below:
\begin{equation*}
a_{ij} = \begin{cases}
	    0,\quad \text{if node $i$ is not a part of link $j$}\\ 
	    +1,\quad \text{if node $i$ is the source to link $j$},\\ 
	    -1, \quad \text{if node $i$ is the dest to link $j$}.
	    \end{cases}
\end{equation*}

$X$ refers to an $L\times F$ matrix with values $X_l^f$ representing the average effective rate of transmission chosen on link $l$ corresponding to flow $f$.

$\mathbf{r}$ refers to an $N\times F$ matrix, which contains the net outwards data rate at each node corresponding to each flow. i.e. 
\begin{equation*}
r_{ij} = \begin{cases}
	    0,\quad \text{if node $i$ is not a source or dest. of flow $j$}\\ 
	    +r_j,\quad \text{if node $i$ is the source to flow $j$},\\ 
	    -r_j, \quad \text{if node $i$ is the destination to flow $j$}.
	    \end{cases}
\end{equation*}

$\mathcal{M}=\{1,2,...,M\}$ refers to the set of all possible \emph{modes} in the network. Each mode $m\in \mathcal{M}$ will take a representation of an $L$ dimensional column vector. The modes set includes an idle state (i.e. a $\underline{0}$) as well.

$\underline{\alpha}$ represents the vector of time fractions given to each of the modes.

$C$ is an $M\times L$ vector, where an element $C_l^m$ represents the capacity of a link $l\in \mathcal{L}$ in mode $m\in \mathcal{M}$. i.e., each column of $C$ is a vector of same dimension as a mode, but link powers are replaced with link capacities.

$P$ refers to an $N\times M$ matrix called \emph{power profile matrix} with values $P_n^m$ representing the amount of power spent by node $n$ in mode $m\in \mathcal{M}$. It corresponds to the power spent on the single active link in the neighborhood of a node as a transmitter, at any time slot.
$\underline{P}^{avg}$ is an $N\times 1$ vector, representing the power availability at each node, on an average.

Note that the variables in the problem are $r_i(\forall i\in\mathcal{F}), X$ and $\underline{\alpha}$ while rest all($A,P,C$ and $\underline{P}^{avg}$) are constants once the network and set of all modes $\mathcal{M}$ are fixed. Also note that, its a pure linear program since the objective can alternately be written as a set of linear inequalities introducing one new variable.

With the set of assumptions we made in the system model, given a network($A$) with average power values($\underline{P}^{avg}$), this problem statement becomes universally applicable for any number of antennas($a$) on nodes, given the set of modes $\mathcal{M}$ and their capacities $C$. But with multiple antennas, we see following issues.\\
1) The link capacity calculations are interdependent (a joint optimization problem) due to interference. We would have a capacity region of points being vectors of link capacity elements.\\
2) The capacity region of such a vector interference channel is still an open problem.\\
3) We need a computationally \emph{feasible} method to decide link capacities in a mode (columns of $C$), atleast as a good achievable rate point. Because, we have to calculate the same for a large number of modes.

\subsection{Link Capacity computation}
Let $\underline{m}=(m_1,m_2,\ldots,m_L)^T$ be the given vector of $L$ powers spent on links. To compute the link capacities, one should solve the problem of waterfilling on each of the links. But, the effective noise that each user sees is a sum of AWG noise and the interference from other users. Assuming each source emits independent Gaussian symbols, we need the covariance matrix of the effective noise, for which we need covariance matrices of each transmitter in the network. This is clearly not known at the time of computation of link capacities. Because, link capacity computation itself is \emph{water flling} i.e. an optimization decision over all transmit covariance matrices possible at all the links.

We propose the following iterative scheme to greedily calculate the capacities of links in the network which are active simultaneously. We continue the iterations till we see the convergence in sum rate of all links in the network. The sum rate increases monotonically after each iteration. 

We use the following notation. $C_i$ is the $i^{th}$ link capacity. $K_i$ is the transmit covariance matrix used on link $i$ while $m_i$ is the power spent on link $i$. $sumrate$ corresponds to the sum of all link capacities active in the network. The function ``waterfill(.)'' performs waterfilling on any specified link and outputs two values namely link capacity and optimum transmit covariance matrix to achieve that capacity under the interference value at that instant. It takes three arguments namely link index at which waterfilling is to be performed, set of all transmit covariance matrices and channels($H_{ij}'$) between link $i$ and every other link $j$ in the network. (i.e., source of a link to the destination of other link --- $H_{ij}' = H_{pq}$ where $p=src(i)$ and $q=dest(j)$).

\begin{algorithm}
   \begin{algorithmic}
    \caption{Iterative Waterfilling algorithm for finding link-capacities in a point-to-point mode}
    \STATE initialize\ \ $sumrate = -\infty, \ \ C_i=0,\ \ K_i=(m_i/a)*I$  
	    $\forall i=\{1,2,\ldots,L\}$ 
    \WHILE{ $|\sum_{i}{C_i} - sumrate| \geq \epsilon$}
      \STATE $sumrate= \sum_{i}{C_i}$
      \FOR{$i=1$ \TO $L$}
	  \IF{ link $i$ is active }
	  \STATE $[K_i, C_i ] = $ waterfill($i; \sigma^2; \{K_j, H_{ji}' \forall j\in \mathcal{L}\} $)
	  \ENDIF
      \ENDFOR
    \ENDWHILE
  \end{algorithmic}
\label{IWF_P2P}
\end{algorithm}

\subsection{Complexity Issues}
\begin{figure}[!htb]
\includegraphics[scale=0.4]{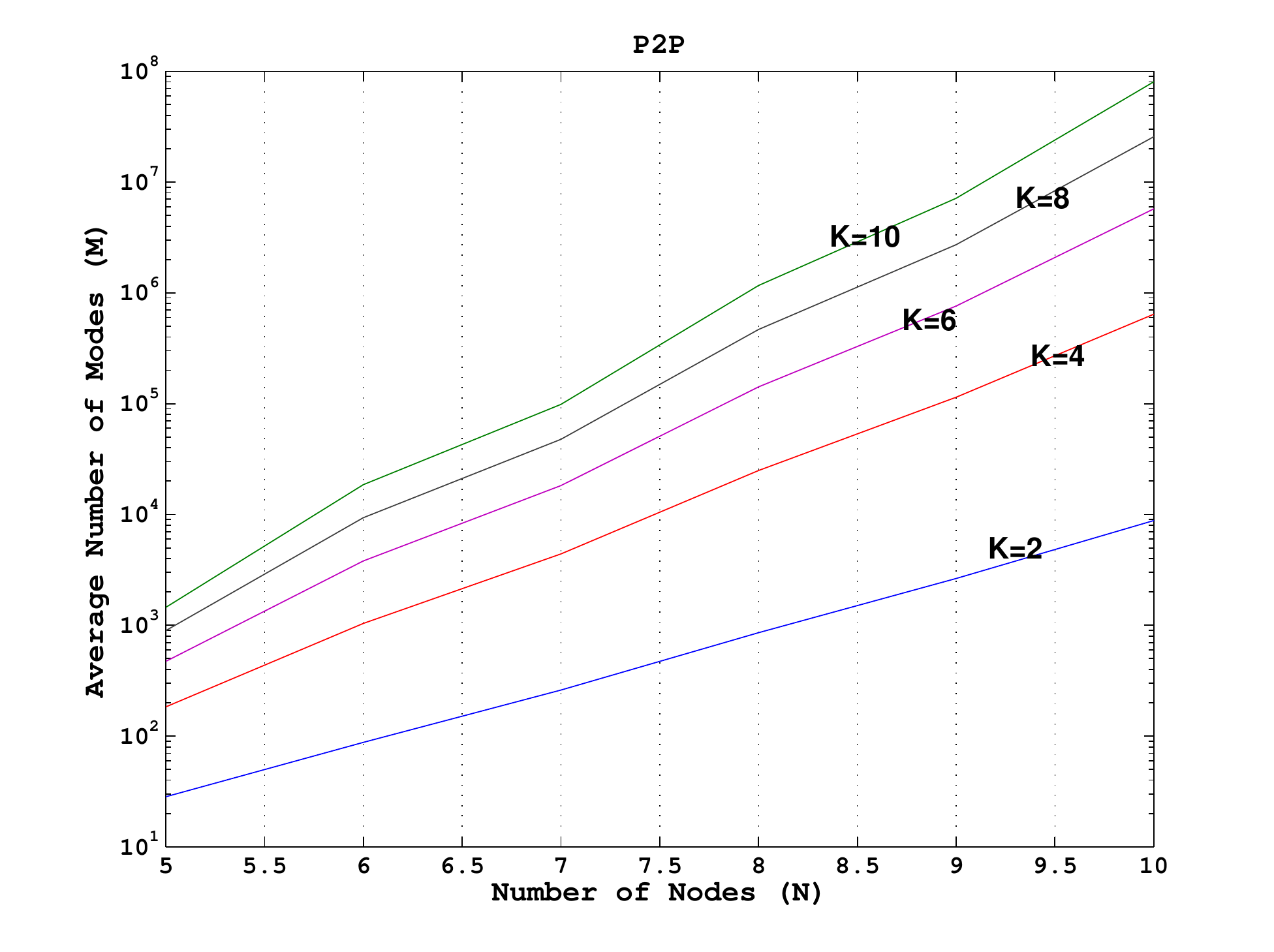} 
\caption{A typical number of modes(M) as a function of number of nodes(N)}
\label{MODES}
\end{figure}

The number of constraints in the problem is $(NF+L+F+1)$ while number of variables is $(F+LF+M)$. If $N$ is taken as the size of the network, $N\leq L\leq N(N-1)$ since graph is assumed to be directed and fully connected, having no self loops. Further $F\leq N(N-1)$.\\
Though one can derive an upper bound on $M$ as $K^L$ (a super exponential number) by arguing that M is less than all possible choices of powerlevels among L links, it can be seen via bruteforce enumerating simulations that the number of possible modes in a network of $N$ nodes is exponential in $N$ (see fig.\ref{MODES}). In anycase, the number of variables in the problem grows enormously with $N$ and is much larger compared to number of constraints in the network. Typically, this number turns out to be more than a million for a network as small as 11 nodes and 4 powerlevels.

Hence, its a prohibitively large problem to be solved directly for networks more than roughly 10 nodes. Note that, we have ignored the calculations required to get link capacities.

\section{The Heuristic Solution}
One definitely needs to have a solution procedure, that is feasible for solving the network, upto atleast reasonable size of the network. \cite{Cao_etal} proposes a heuristic column generation method to solve this problem and is claimed to solve networks upto approximately 30 nodes. \cite{VJ_etal} applies the similar thing with slight variation, on the networks of energy harvesting sensor nodes. We see that we can still solve the problem upto almost the same level, even when the network has multiple antennas and hence the complexity is much larger.

The solution method proceeds as follows. First it can be seen by using the analysis of number of basic variables that, no more than $N+L+1$ number of modes are necessary to obtain the optimal basic feasible solution. Then as per the \emph{column generation} procedure, solution starts by considering a smaller problem called \emph{Master Problem} with just $N+L+1$ randomly chosen mode variables($\underline{\alpha}'$) from $\underline{\alpha}$, rest all being the same to the original optimization problem in \eqref{The_LPP}. 

It also considers a \emph{Sub Problem} that chooses a new mode called a \emph{good mode} from the remaining modes, considering which we would improve the performance of the master problem. \\
\textbf{Sub Problem:}
\begin{align*}
 \max_{m\in \mathcal{M}\text{\textbackslash}\mathcal{M'}}{ \theta(m) \triangleq \underline{u}^T\underline{C}_m - \underline{v}^T\underline{P}^m - \beta} \\
subject~~to:\\ \theta(m) \geq 0.
\end{align*}
Such a mode will replace an already chosen mode variable which is given zero scheduling time in the optimal solution of \emph{Master} problem. This starts next iteration, in which master problem attains an improved solution. \\
The same after iterating for sufficient number of iterations, converges to a (sub)optimal solution and an (sub)optimal choice of set of $N+L+1$ modes from $\mathcal{M}$.

In this procedure, sub-problem is the stage, which requires an optimum choice of variable by searching exhaustively among auxiliary function values evaluated at $M-(N+L+1)$ modes, which is again computationally infeasible. This can be avoided by using a heuristic algorithm for sub-problem, from \cite{Cao_etal} and \cite{VJ_etal}. This comes at the cost of sub-optimality to the solution. But it was shown via simulation examples that, the solution that we see via this heuristic, is close to the optimal solution.

The heuristic algorithm for MHWN with MIMO extended from its single antenna version is presented below as Algorithm \ref{Heuristic}, in a ready to implement algorithmic form. In this, $nextlevel(x)$ is the notation used for denoting the power value which is least among the power values that are greater than $x$ from power level set. And $interferers(l)$ is the notation used to denote the \emph{set} of all the neighbouring links to link $l$ that \emph{share} a node with link $l$.

\begin{algorithm}
{\footnotesize
 \begin{algorithmic}
  \caption{The Heuristic Algorithm to obtain an efficient \emph{sub-optimal} solution to the \emph{sub-problem}}
  \label{Heuristic}
    \STATE Initialize mode vector $\mathbf{goodmode}=\underline{0}$,\\ $\mathbf{lasttheta}=\theta(\mathbf{goodmode})$, $\mathbf{AllowedSet}=\mathcal{L}$\vspace{2mm}
    \WHILE{$\mathbf{AllowedSet}\neq \phi$}
    \FOR{$i=1$ \TO $L$}
      \IF{ $i \in \mathbf{AllowedSet}$ }
	\STATE $\underline{m} = \mathbf{goodmode}$ 
	\STATE $m_i=nextlevel(m_i)$ 
	\STATE $\theta_i = \theta(\underline{m})$ 
      \ENDIF
    \ENDFOR \vspace{2mm}
	\STATE Let $\mathbf{besttheta} \triangleq \max_{i}{\theta_i}$ ~~\&
	\STATE Let the $\mathbf{bestlink}~l \triangleq arg~\max_{i}{\theta_i}$
	\IF{$\mathbf{besttheta} \leq \mathbf{lasttheta}$}
	  \STATE break the loop.
	\ELSE
	  \STATE $\mathbf{lasttheta} = \mathbf{besttheta}$
	  \STATE $\underline{m}=\mathbf{goodmode}$
	  \IF{$m_l=0$}
	    \STATE $\mathbf{AllowedSet}=\mathbf{AllowedSet}\backslash interferers(l) $
	  \ENDIF
	  \STATE $m_l = nextlevel(m_l)$
	  \STATE $\mathbf{goodmode}=\underline{m}$
	\ENDIF
    \ENDWHILE
    \STATE Declare $\mathbf{goodmode}$ as the solution to sub problem
 \end{algorithmic}
}
\end{algorithm}

Since we solve the problem \eqref{The_LPP} with a heuristic solution at the sub problem level of column generation, we call this method as \emph{Heuristic Column Generation} (HCG). It should be noted that, this solution is sub-optimal for the sub-problem and hence the final solution that we obtain by using this is sub-optimal. 

Further, it can be predicted and also observed in simulations presented in next section that, the final solution (maximum fairness) varies over a range of values, depending upon the initially selected random set of modes. This issue was not discussed much in the earlier papers using this HCG. Hence we in this paper, propose to choose the best value among solutions achieved by solving the problem for multiple number of initial points, analogous to the traditional tabu search methods used for non-convex optimization. 

In next chapter, we solve the problem for several networks with multiple antennas and get the final solutions. We first show for appropriate networks that the HCG gives a sub-optimal solution but is very close to the optimal solution. Then we see that we can solve the problem upto 30 nodes, almost similar to the single antennas case, as claimed in earlier papers. Further, we show that the MIMO does improve the final performance of the network and further that the gain is linear with the number of antennas($a$) on nodes.

\section{simulation results}
\begin{figure}[!ht]
\centering
\includegraphics[scale=0.5]{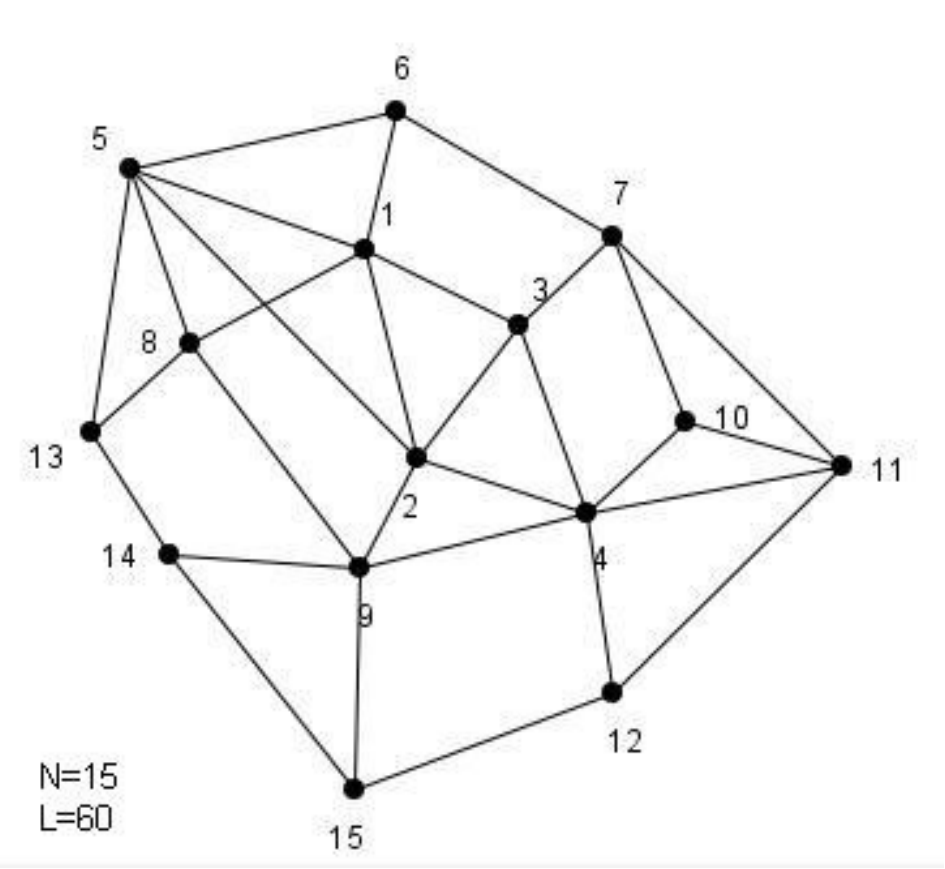}
\caption{A network of 15 nodes}
\label{N15graph}
\end{figure}

Consider a network of 15 nodes as shown in fig-\ref{N15graph}.
Consider the following network parameters:\\
Number of Nodes $N=15$\\
Number of Directed Links $L=60$\\
Number of antennas on each node $a=5$\\
Power levels used by each node = $[0,4]$;\\
Channels are generated as $L$ number of $5\times 5$ matrices of unit variance, circularly Gaussian complex random numbers.\\
source,destination pairs(flows) are =  (7,13), (10,5), (11,8)\\
The rate demands from the flows =  $[10, 15, 20]$units\\
Nodes Avg power availability = 3 units, uniform to all.\\
(Note: In this scenario, we have approximately 0.5 million modes allowed in the network)

\noindent By formulating the problem \eqref{The_LPP} and solving it directly (optimally), we get the following solutions.
Allowed user rates after solving the problem are:

\begingroup \centering
$r_1 = 5.971$ units,  $r_2 = 8.957$ units\\
$r_3 = 11.942$ units\\
Hence the optimum fairness that can be achieved is:\\
$\lambda = 0.5971$\\
\endgroup

\begin{figure}[!ht]
 \centering
  \includegraphics[scale=0.4]{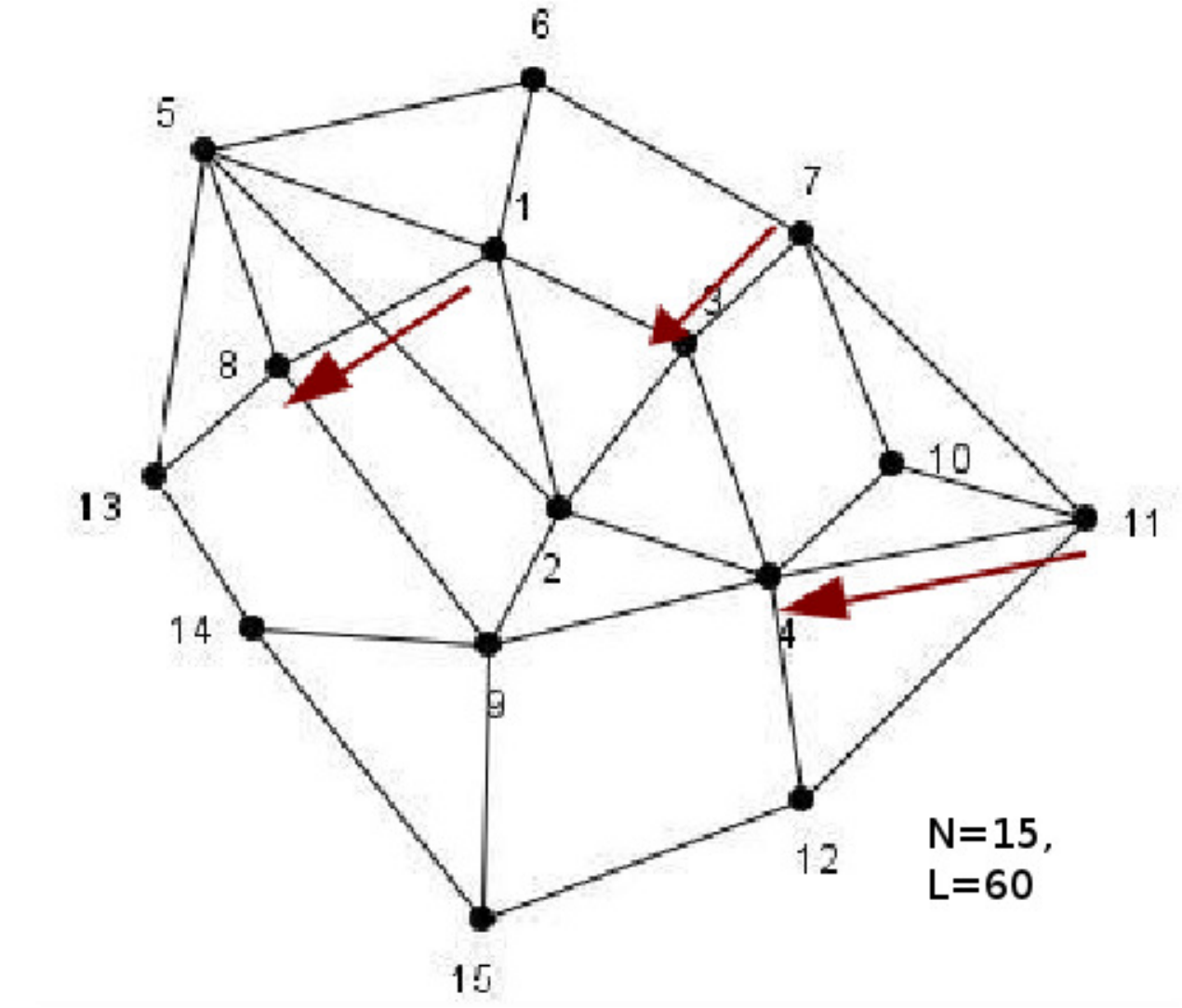}
\caption{A dominant mode in the network to achieve optimal throughput}
 \label{Dom_Modes_N15}
\end{figure}

A dominant mode which is given fairly good amount of time fraction is shown in fig-\ref{Dom_Modes_N15} with numbered bold arrows.

\begin{figure}[!ht]
 \centering
  \includegraphics[scale=0.5]{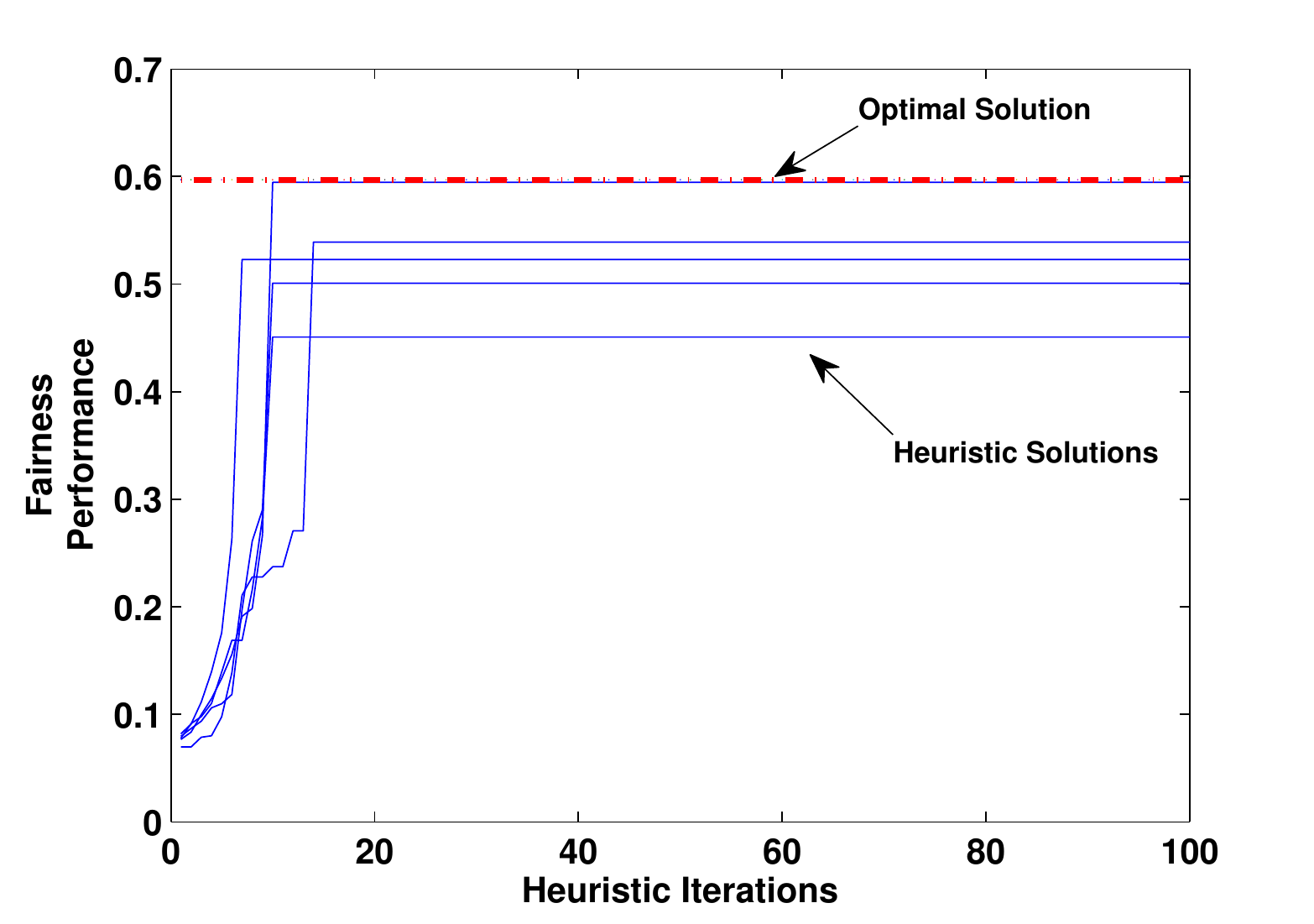}
\caption{Fairness Values achieved using Heuristic Algm, Compared with Optimal for the network of 15 nodes}
 \label{Heuristic_N15}
\end{figure}

Now we use Heuristic algorithm to solve the problem and see the relative performance. This is shown in fig-\ref{Heuristic_N15}. Note that we use multiple initial points. we get the following solutions.\\
Allowed user rates after solving the problem are:

\begingroup \centering
$r_1 = 5.8$ units,  $r_2 = 8.7$ units\\
$r_3 = 11.6$ units\\
Hence the optimum fairness that can be achieved is:\\
$\lambda = 0.58$\\
\endgroup

\noindent Next, we consider a network of 30 nodes and the following parameters.

\begin{figure}[!ht]
\centering
\includegraphics[scale=0.27]{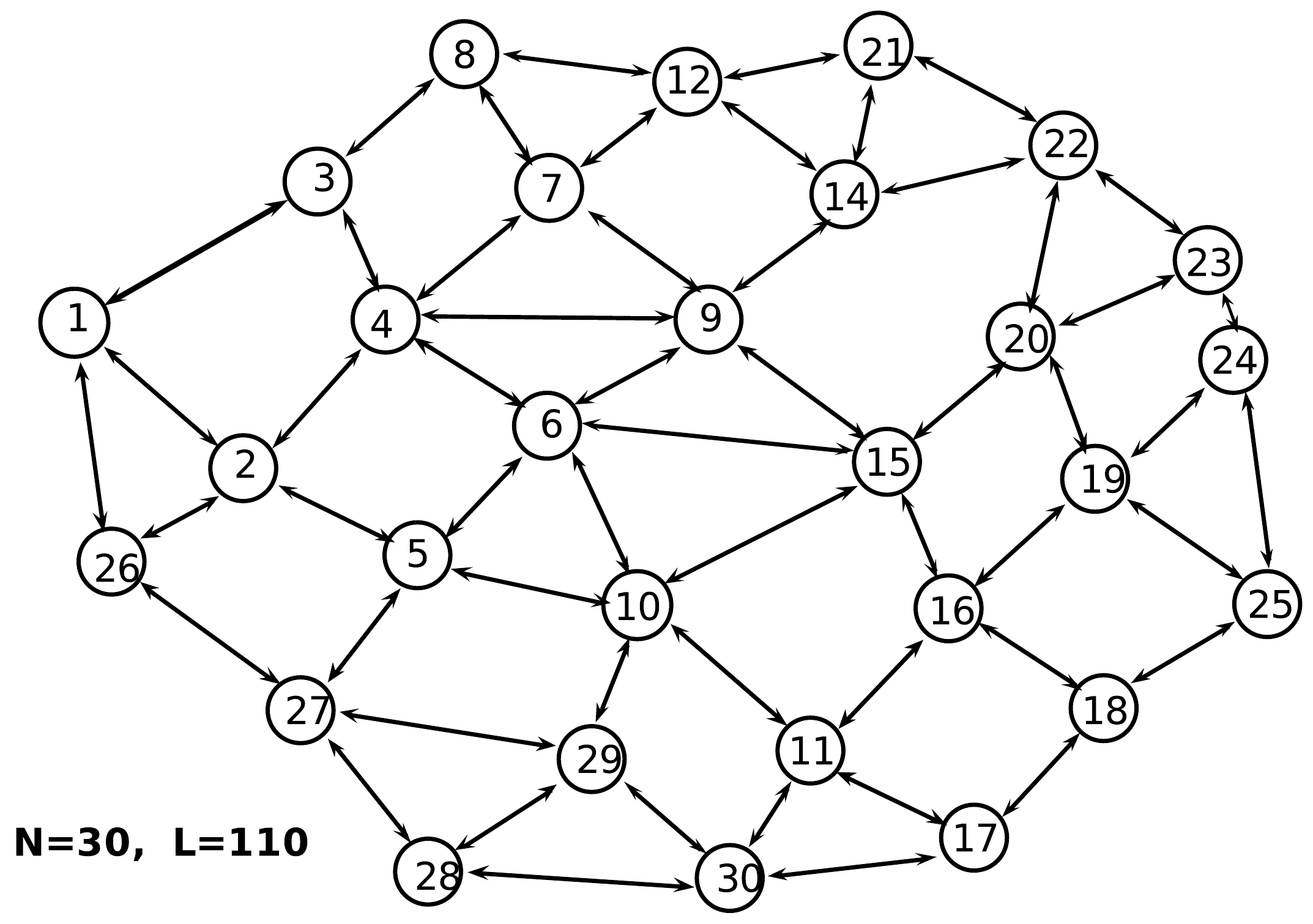}
\caption{A network of 30 nodes}
\label{N30graph}
\end{figure}

\noindent Number of Nodes $N=30$\\
Number of Directed Links $L= 110$\\
Number of antennas on each node $a=4$\\
Source,destination pairs(flows) are: \\ 
=  (2,30), (5,10), (1,30), (4,6)\\
The rate demands from the flows =  $[10, 50, 20, 60]$ units\\
Nodes Avg power availability = 3 units, uniform to all.

\begin{figure}[!ht]
 \centering
  \includegraphics[scale=0.5]{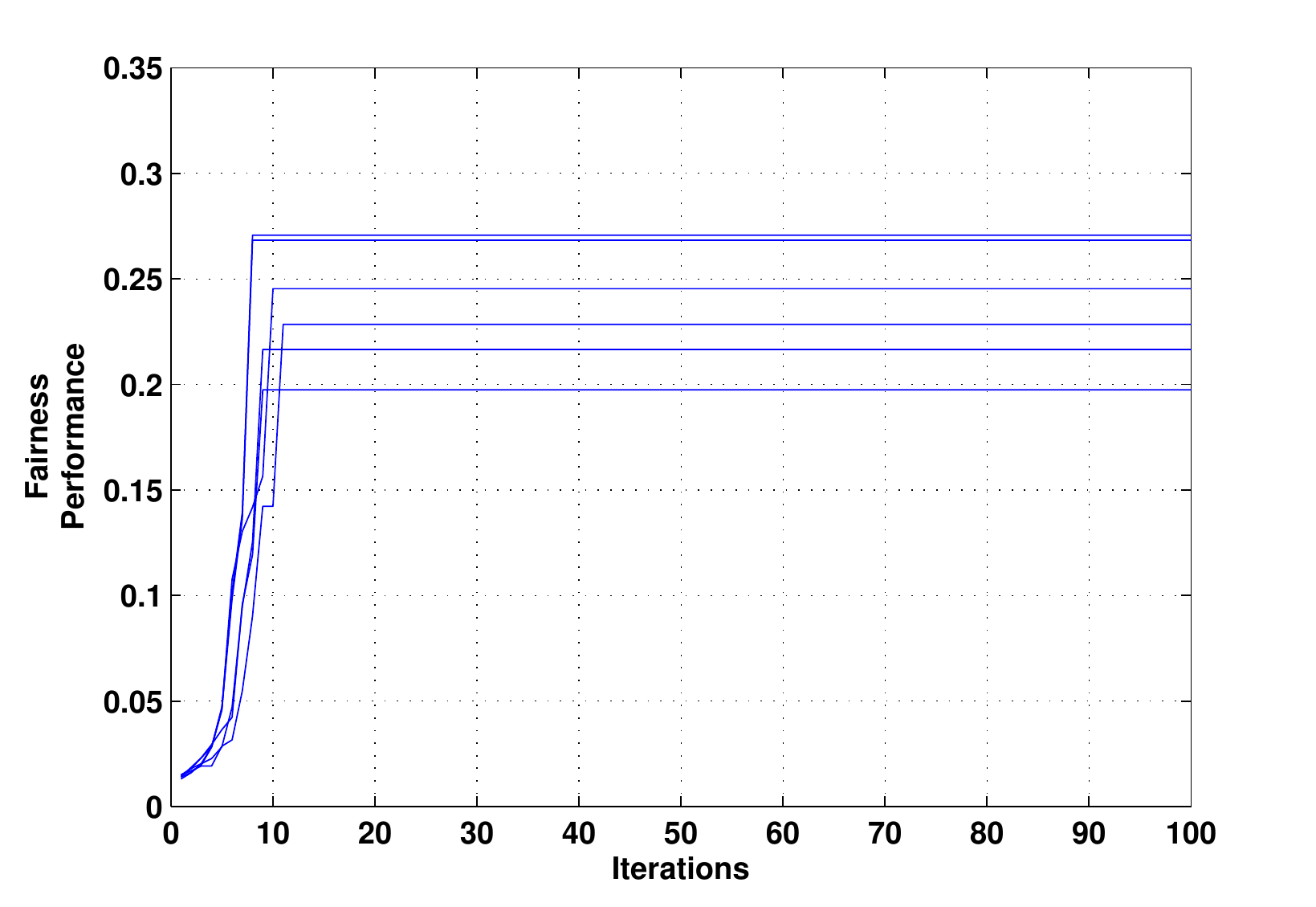}
\caption{Heuristic algorithmic performance for network of N=30}
 \label{Heuristic_N30}
\end{figure}

\noindent Formulating the problem \eqref{The_LPP} and solving it using Heuristic for multiple initial points, we see the following results fig-\ref{Heuristic_N30}. Note that, this problem cannot be solved directly and hence the optimal performance achievable is not known.

\vspace{2mm}
\noindent Allowed user rates after solving the problem are: \\
\begingroup \centering
$ r_1 = 2.7069$ units,  $ r_2 = 13.5345$ units\\
$ r_3 = 5.4138$ units,  $ r_4 = 16.2414$  units\\
Hence the optimum fairness that can be achieved is:\\
$ \lambda = 0.2707$\\
\endgroup

\noindent So we demonstrated that proposed system model with JRSP problem is solvable upto reasonable size of the network and that the extended \emph{heuristic} solution gives close enough solutions to the optimal values. We now show the performance gain achieved using multiple antennas in the system.

\begin{figure}[!ht]
 \centering
  \includegraphics[scale=0.5]{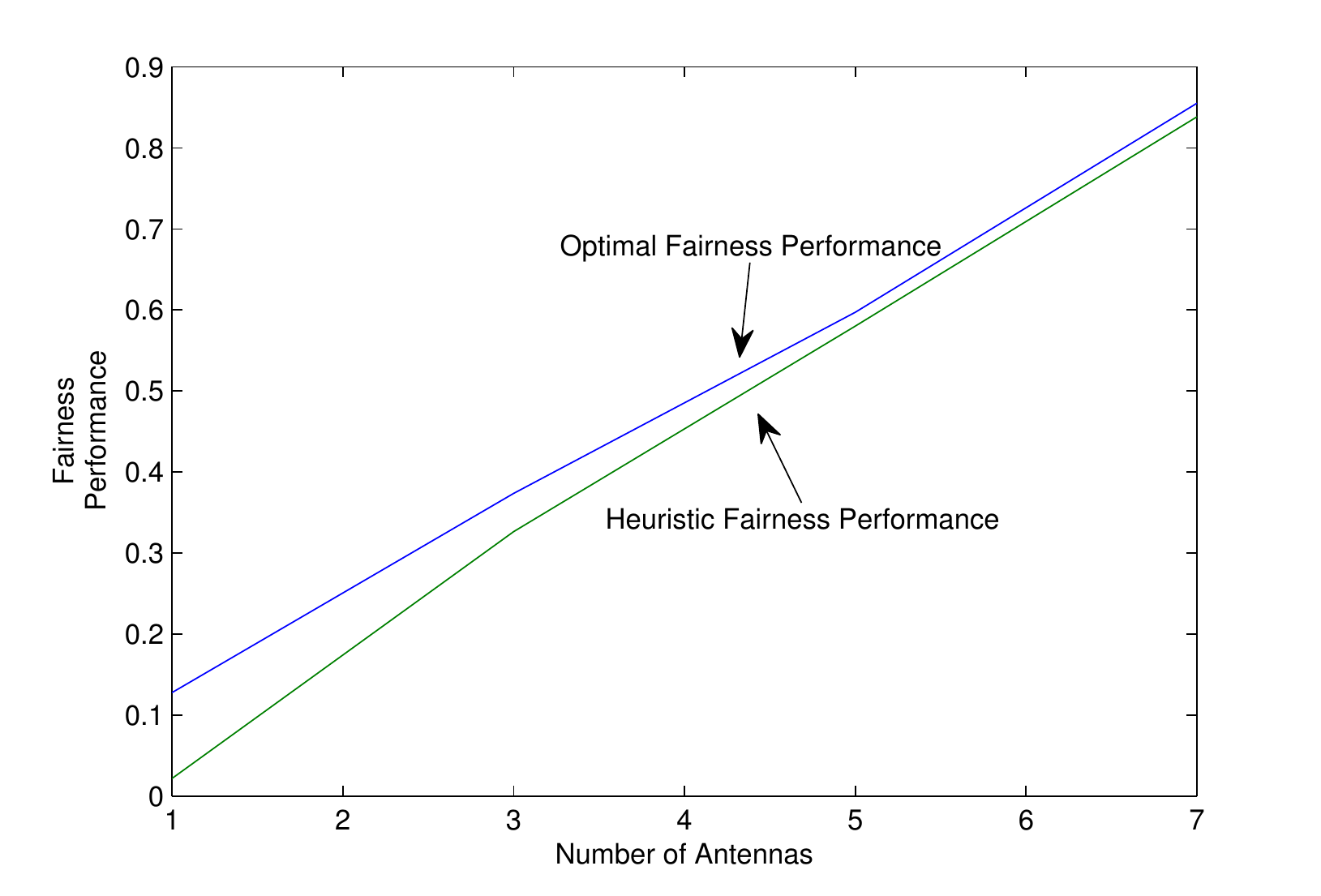}
\caption{MIMO gain in Optimal and Heuristic algorithmic performance of a network with 15 nodes}
 \label{Heu_MIMO_gain_N15}
\end{figure}

We consider the same 15 node network and the problem shown in fig-\ref{N15graph}. We vary the number of antennas on each node($a$) from one to ten and see how the optimal performance (fairness) varies. Note that the channel matrices observe a change in dimensions as well as values, for each case. We then consider the same network problem but solve it using the Heuristic column generation. We vary the number of antennas ($a$) from one to ten and see the performance (fairness) variation. Both these are shown in fig-\ref{Heu_MIMO_gain_N15}
We see that both optimal as well as Heuristic algorithmic performances improve linearly as we increase the number of antennas on nodes.

\section{Conclusion}
We have proposed a feasible system model for multihop wireless networks with multiple antennas extended from the MHWN with single antennas. We have shown that the problem is solvable upto reasonale size of the network almost similar to the network with single antennas. We then have shown that the performance gain that we achieve using multiple number of antennas on nodes is linear.

\end{document}